\documentclass[11pt]{article}
\usepackage{amsfonts, latexsym, amssymb}
\makeatletter
\def\bee{\begin{equation}}
\def\eee{\end{equation}}

\def \al{\alpha}
\def \be{\beta}

\def \0{{\theta_0}}
\def \1{{\theta_1}}

\def \underset#1#2{\mathrel{\mathop{#2}\limits_{#1}}}
\@addtoreset{equation}{section}

\def\bm{\left(\begin{array}{cc}}
\def\em{\end{array}\right)}

\newtheorem{remark}{Remark}[section]

\def\text#1{\mbox{#1}}

\def\lim{\mathop{\mbox{lim}}\limits}

\makeatother
\begin{document}

\title{Some Examples of $RS^2_3(3)$-Transformations of Ranks $5$ and $6$
as the Higher Order Transformations for the Hypergeometric Function}
\author{F.~V.~Andreev and A.~V.~Kitaev
\thanks{E-mail: andreev@math.ksu.edu;  akitaev@pdmi.ras.ru}\\
Steklov Mathematical Institute, Fontanka 27, St.Petersburg,
191011, Russia*\\
Kansas State University, Department of Mathematics,\\ 
Manhattan, KS, 66506, USA}

\date{December 20, 2000}
\maketitle

\begin{abstract}
A combination of rational mappings and Schlesinger transformations
for a matrix form of the hypergeometric equation  
is used to construct higher order transformations for the Gauss 
hypergeometric function.\vspace{24pt}
\vspace{24pt}

\noindent 
Short title: $RS$-Transformations for the Hypergeometric Function
\end{abstract}
\maketitle
\newpage

\setcounter{page}2
\section{Introduction}
 \label{Introduction}
This is a continuation of our series of works devoted to the
so-called $RS$-transformations for linear matrix ODEs.
The notation $RS^n_p(q)$ means $RS$-transformation of
$n\times n$ matrix linear ODE with $q$ singular points into
another linear ODE with $p\geq q$ singular points. The abbreviation
$RS$ is related with the fact that these transformations are
compositions of rational transformations of arguments
of the linear ODEs with the Schlesinger transformations of their
solutions. The rank of the $RS$-transformation is by definition the order 
(a number of preimages) of its rational transformation.
It would be very interesting to study these 
transformations for $n>2$, however in this, as well as in our
previous works, we restrict our consideration to the case 
of $2\times2$ matrices and sometimes 
omit the superscript $n=2$ to simplify the notation. 

In the work \cite{K1} it was shown that transformation $RS_4(4)$
generates quadratic transformation for the sixth Painlev\'e
equation, in \cite{K2} a general notion of the $RS$-transformation
for the so-called special functions of the isomonodromy type \cite{K3}
were introduced and application to the construction of the algebraic solutions
to the sixth Painlev\'e equation was discussed, in \cite{AK} we
classified all $RS_4(3)$ transformations of the ranks $\leq4$.

In this work we show that some higher order transformations of the
Gauss hypergeometric function can be obtained by means of the 
$RS_3(3)$-transformations. More precizely, $RS_3^2(3)$-transformations of the 
rank $r$ 
``generate'' so-called transformations of order $r$ for the hypergeometric 
function.
Actually, we believe that one can obtain by this
method all the ``seed'' transformations of the higher order for the 
hypergeometric
function, i.e., few transformations from which all other higher order 
transformations can be derived via already known transformations for the 
hypergeometric function. 

To the best of our knowledge explicit examples of the higher order
transformations for the hypergeometric functions, which are not
combinations of the quadratic and cubic transformations, are unknown.
The main goal of this paper is to present some explicit examples of such 
transformations of the orders $5$ and $6$. We hope to release soon a
complete classification of the seed transformations of the orders $4-6$.
  
It is written in the book \cite{BE}, with the reference to the works of
Goursat dated back to 1881 and 1938, that transformations for the hypergeometric 
function of the orders different from $2$, $3$, $4$ and $6$
exist only in those cases when it is algebraic, i.e., it corresponds to some 
case in
the famous H.~A.~Schwartz table of 1873. Actually, this statement is not true
since for the hypergeometric function corresponding to the Schwartz parameters
$\left(1/2,1/2,\nu\right)$, where $\nu$ is an arbitrary number, one can apply
proper quadratic or cubic transformations arbitrary number of times. 
This statement is also not true if we exclude  
higher order transformations related with iterations of the transformations of 
the 
orders $\leq6$. In fact, in the last Section of this paper, we show that there 
is 
an eighth order transformation relating hypergeometric functions whose 
parameters
are not included into the Schwartz table. We know also some other examples
of this kind. 
 
The paper is organized as follows.
In Section \ref{sec:hyper} we,
following the work \cite{J}, write the Euler equation for the Gauss 
hypergeometric
function in the matrix form. Actually solutions of this matrix equation can
be parameterized in terms of the hypergeometric functions in a variety of ways
(recall 24 Kummer's solutions for the Euler equation). Clearly, for different
transformations it is better to choose those representation of solutions in 
terms of
the hypergeometric series in  which these transformations look simpler. We are 
not
doing it here and using the same form of the fundamental solution for all 
transformations under consideration. 
However our formulas can be easily adopted for any other representations of
the fundamental solutions. The only difference will be in the right-hand
factor in the corresponding $RS$-transformations, which is denoted for all
transformations as the matrix $V$.
In Section \ref{sec:2} we explain basic principles for classification of
the $RS_3^2(3)$-transformations and corresponding notation.
In Section \ref{sec:qucu} a brief account of the quadratic and cubic
transformations is given.
The following two Sections \ref{sec:3} and \ref{sec:6} are devoted to 
the examples of the $RS$-transformations of the ranks $5$ and $6$, respectively.
Finally, in Section \ref{sec:8} we discuss an example of the $RS$-transformation
of the rank $8$ mentioned above. 
  
{\bf Acknowledgment} The work was stimulated by the lecture \cite{B}
given by B.~C.~Berndt at NATO ASI Workshop ``Special Functions 2000'' 
and subsequent conversations with Richard Askey. The authors are grateful to the 
organizers of the NATO ASI Workshop M.~Ismail, J.~Bustoz, V.~Spiridonov, 
and S.~Suslov for the invitation and financial support.
During preparation of this work A.~V.~K. was supported
by Alexander von Humboldt-Stiftung and hosted by Universit\"at GH Paderborn. 
\section{Hypergeometric Equation}
 \label{sec:hyper}
Consider the following matrix form of the hypergeometric equation,
\begin{equation}
 \label{eq:PHI}
\frac{d\Phi}{d\lambda}=\left(\frac{A}{\lambda}+\frac{B}{\lambda-1}\right)\Phi,
\end{equation}
where we, following \cite{J}, parameterize the matrices $A$ and $B$ 
by three complex numbers, $\alpha$, $\beta$, and $\delta$,
$$
A=\left(\begin{array}{cc}
-\frac{(\alpha+\beta)(1-\delta)+2\alpha\beta}{2\beta-2\alpha}&
\frac{\beta(\beta+1-\delta)}{\beta-\alpha}\\
-\frac{\alpha(\alpha+1-\delta)}{\beta-\alpha}&
\frac{(\alpha+\beta)(1-\delta)+2\alpha\beta}{2\beta-2\alpha}
\end{array}\right),\quad
B=-A+\frac{\beta-\alpha}2\sigma_3.
$$
Explicit formula for the fundamental solution $\Phi$ in terms of 
the Gau{\ss} hypergeometric functions, as well as asymptotic behavior
of $\Phi$ in the neighborhood of the singular points, 
can be found in \cite{J}. We'll use these formulas and therefore
present them below for reference.
\begin{eqnarray}
&
\Phi(\lambda;\theta_0,\theta_1,\theta_\infty)
=&\label{eq:PHI_HYPER}\\
&\bm F(\al,\al-\delta+1;\al-\be;\frac 1\lambda) &
\frac{\be(\be-\delta+1)}{(\be-\al)(\be-\al+1)}\frac1\lambda\\
&\times F(\be+1,\be-\delta+2;\be-\al+2;\frac1\lambda)\\
\frac{\al(\al-\delta+1)}{(\al-\be)(\al-\be+1)}
\frac1\lambda 
&
F(\be,\be-\delta+1;\be-\al;\frac1\lambda)\\
\times F(\al+1,\al-\delta+2;\al-\be+2;\frac1\lambda)&
\em\times &\nonumber\\
&\times\lambda^{-\bm \al & \\ & \be\em}
\lambda^{\frac{\delta-1}2} 
(\lambda-1)^{\frac{\al+\beta-\delta+1}2},&\nonumber
\end{eqnarray}
where $F(\cdot,\cdot;\cdot;\cdot)$ is the Gauss hypergeometric function 
\cite{BE}
We define parameters of the formal monodromy, $\theta_k$,
as follows:
\begin{equation}
 \label{eq:theta}
\theta_0=1-\delta,\quad \theta_1=-\alpha-\beta+\delta-1,\quad \theta_\infty
=\alpha-\beta.
\end{equation}
The inverse formulas read:
$$
\alpha =\frac12(-\theta_0-\theta_1+\theta_\infty),
\quad
\beta =\frac12(-\theta_0-\theta_1-\theta_\infty),
\quad 
\delta=1-\theta_0.
$$
The function $\Phi(\lambda;\theta_0,\theta_1,\theta_\infty)$ has the
following asymptotic behavior at singular points,
\begin{eqnarray}
\Phi(\lambda;\theta_0,\theta_1,\theta_\infty)&\underset{\lambda\to0}=&
G_0(\alpha,\beta,\delta)(I+{\cal O}(\lambda))
\lambda^{\frac{\theta_0}2\sigma_3}C_0(\alpha,\beta,\delta)
\label{as:0}\\
&\underset{\lambda\to1}=&
G_1(\alpha,\beta,\delta)(I+{\cal O}(\lambda-1))
(\lambda-1)^{\frac{\theta_1}2\sigma_3}C_1(\alpha,\beta,\delta)
\label{as:1}\\
&\underset{\lambda\to\infty}=&
\left(I+{\cal O}\left(\frac1\lambda\right)\right)
\left(\frac1\lambda\right)^{\frac{\theta_\infty}2\sigma_3},
\label{as:infty}
\end{eqnarray}
where $I={\rm diag}\{1,1\}$ and $\sigma_3={\rm diag}\{1,-1\}$ are 
$2\times2$ diagonal matrices,
$$
G_0(\alpha,\beta,\delta)=\frac1{\beta-\alpha}
\left(\begin{array}{cc}
\beta-\delta+1&\beta\\
\alpha-\delta+1&\alpha
\end{array}\right),\qquad
G_1(\alpha,\beta,\delta)=\frac1{\beta-\alpha}
\left(\begin{array}{cc}
1&\beta(\beta-\delta+1)\\
1&\alpha(\alpha-\delta+1)
\end{array}\right),
$$
$$
C_0(\alpha,\beta,\delta)=
e^{\frac{i\pi}2(\alpha+\beta-\delta+1)}\!\left(\begin{array}{cc}
e^{-\pi i(\alpha-\delta+1)}
\frac{\Gamma(\delta-1)\Gamma(\alpha-\beta+1)}
{\Gamma(\delta-\beta)\Gamma(\alpha)}&-e^{-\pi i(\beta-\delta+1)}
\frac{\Gamma(\delta-1)\Gamma(\beta-\alpha+1)}
{\Gamma(\delta-\alpha)\Gamma(\beta)}\\
e^{-\pi i\alpha}
\frac{\Gamma(1-\delta)\Gamma(\alpha-\beta+1)}
{\Gamma(1-\beta)\Gamma(\alpha-\delta+1)}&-e^{-\pi i\beta}
\frac{\Gamma(1-\delta)\Gamma(\beta-\alpha+1)}
{\Gamma(1-\alpha)\Gamma(\beta-\delta+1)}
\end{array}\right)\!,
$$
$$
C_1(\alpha,\beta,\delta)=
\left(\begin{array}{cc}
-\frac{\Gamma(\alpha+\beta-\delta+1)\Gamma(\alpha-\beta+1)}
{\Gamma(\alpha-\delta+1)\Gamma(\alpha)}&
\frac{\Gamma(\alpha+\beta-\delta+1)\Gamma(\beta-\alpha+1)}
{\Gamma(\beta-\delta+1)\Gamma(\beta)}\\
-\frac{\Gamma(-\alpha-\beta+\delta-1)\Gamma(\alpha-\beta+1)}
{\Gamma(1-\beta)\Gamma(\delta-\beta)}&
\frac{\Gamma(-\alpha-\beta+\delta-1)\Gamma(\beta-\alpha+1)}
{\Gamma(1-\alpha)\Gamma(\delta-\alpha)}
\end{array}\right).
$$
\section{$RS$-transformations}
 \label{sec:2}
For any matrix linear ODE, e.g., the hypergeometric
equation from the previous Section, one can consider, as we call them, 
$RS$-transformations. These transformations map a given ODE
into another ODE and consist of transformations of two types,
which we call for brevity $R$- and $S$- transformations.
The first one ($R$-) is a rational transformation of the argument
of the ODE, and the name $S$-transformation stands for
the subsequent Schlesinger transformation of the solution. In some cases
it is possible to construct auto-$RS$-transformations which
map a given ODE to itself. In this paper we deal with 
auto-$RS$-transformations for Eq.~(\ref{eq:PHI}).
The idea is
to find a proper $R$-transformation for a given ODE so that,
after a special choice of the parameters of formal monodromy,
the parameters $\theta_0$, $\theta_1$, and $\theta_\infty$
of Section \ref{sec:hyper}, one gets in the transformed ODE
$3+N$ singular points, where $N$ of them are apparent.
Therefore, they can be removed via suitable
$S$-transformations. 

It is clear that as far as a proper $R$-transformation is found
one can always construct $S$-transformation which brings 
$R$-transformed Eq.~(\ref{eq:PHI}) into its original form.
Therefore, it is very important to classify all proper
$R$-transformations. For this purpose, it is useful to introduce a notion
of the rank, $r$, of the $RS$-transformation, which is equal to the order 
(a number of preimages) of the corresponding rational function. 
The rational transformations can be enumerated in terms of the triples of 
partitions of $r$ which correspond to the multiplicities of the preimages 
of the points $0$, $1$, and $\infty$. For example, we denote as 
$R(3+2|2+2+1|4+1)$
a rational transformation, $\lambda=R(\lambda_1)$, such that 
$\lambda=0$ has two preimages of multiplicities
$3$ and $2$, $\lambda=1$ has three preimages of multiplicities
$2$, $2$, and $1$, and $\lambda=\infty$ has two preimages of
multiplicities $4$ and $1$. Obviously,  not for any triple of
the partitions one can construct corresponding $R$-transformation,
at the same time many triples for which $R$-transformation exists
are useless for the construction of the $RS$-transformations, since
they do not generate a proper number of the apparent singularities. 
For low orders of $r\leq6$ it is easy (relatively) to construct all possible
$R$-transformations explicitly, whilst for $r\geq7$ one can make
in many cases only theoretical conclusion concerning existence of
such transformation, whilst its explicit form could remain unknown.
As far as 
a proper $R$-transformation is constructed one can make, sometimes,
few different choices of the parameters of formal monodromy to generate
a desired number of the apparent singularities and therefore one arrives
to different $RS$-transformations related with the given $R$-transformation. 
To reflect this we use the following notation for $RS$-transformations
for the Fuchsian ODEs with three singular points,
\par
$$
RS_k^n\left(
\begin{tabular}{c|c|c}
$\theta_0$&$\theta_1$&$\theta_\infty$\\
$p_0(r)$&$p_1(r)$&$p_\infty(r)$\\
\end{tabular}\right),
$$
where $n$ stands for matrix dimension of the ODE under consideration,
$k$ denotes a number of non-apparent singularities in the $R$-transformed
ODE, and
$p_l(r)$, $l=0$, $1$, $\infty$, are the corresponding partitions of $r$.
When we refer to all $RS$-transformations which are associated with
the same $R$-transformation, we denote them as
$RS_k^n(p_0(r)|p_1(r)|p_\infty(r))$, finally, we denote just
$RS^n_k(m)$ the whole set of $RS$-transformations of the Fuchsian
ODE with $m$ singular points to the one with $k$ singular points.  

It is clear that each $RS$-transformation generates some higher
order transformations for the hypergeometric function. On
the other hand, many quadratic and cubic transformations were
obtained by a variety of different ad hoc methods, in particular,
related with the Euler hypergeometric equation. The form of 
these transformations, all of them are linear with respect to
the hypergeometric function, makes natural to suppose that
all of them can be derived from
$RS$-transformations of the rank $r\leq3$. It should
be noted that each $RS$-transformation is only a starting point
for construction of a number of the higher order transformations 
for the hypergeometric function due to the fact that for this
function there is a list of the certain transformations and 
it is not a singlevalued function in $\mathbb{C}$. For example, consider
$RS$-transformations of Eq.~(\ref{eq:PHI}) of the rank $1$. 
They actually coincide with the corresponding $R$-transformations of 
order $1$, since they are just fractional-linear transformations of  
Eq.~(\ref{eq:PHI}) generated by the following two elementary mappings:
$\lambda\to1/\lambda$ and $\lambda\to1-\lambda$, which
produce no any additional singularities in Eq.~(\ref{eq:PHI})
and therefore there is no need to apply any $S$-transformations.
These $R$-/ $RS$-transformations of the rank $1$ clearly ``explain'' 
the relations between the hypergeometric series representing
Kummers solutions, however
additional efforts should be cast to get these relations explicitly.
Even if one knows explicit form of the two ``generating'' relations,
to get the whole list, which is very important for analytical
continuation of the solutions, is not absolutely straightforward.
Looking at the quadratic transformations (see \cite{BE}) one can easily notice,
that actually there is only one seed transformation, whilst all
the others can be obtained via the linear transformations, relations
between the Kummer solutions, the Gauss relations between adjacent
hypergeometric functions, and taking the inverse of the $R$-transformations.
The number of cubic transformations is also can be reduced to few
seed transformations. Although, we haven't checked yet the whole Goursat
list it is very likely that the seed transformations are exactly those
which are generated by the corresponding $RS$-transformations of the rank $3$.  

Having in mind the discussion made in the previous paragraph, it seems
reasonable to suggest the following scheme for classification of
$RS_3^2(3)$-transformations which generate the seed higher order transformations 
for the hypergeometric function. 

{\it We call two $R$-transformations $\lambda=R(\lambda_1)$ and
$\hat\lambda=\hat R(\hat\lambda_1)$ equivalent if there
exist two fractional-linear transformations $f$ and $f_1$
such that $\hat\lambda=f(\lambda)$ and $\hat\lambda_1=f_1(\lambda_1)$}.\\
It is convenient to think about
$RS$-transformations in terms of the parameters of formal monodromy,
i.e., in terms of the triples $(\theta_0,\theta_1,\theta_\infty)$.
These parameters, completely define Eq.~(\ref{eq:PHI}) and, via 
Eq.~(\ref{eq:PHI_HYPER}), corresponding hypergeometric functions.
Since with each $R$-transformation one can associate an infinite
number of $RS$-transformations, via application of the auto-$S$-transformations
to Eq.~(\ref{eq:PHI}), it is reasonable to classify $RS$- transformations modulo 
$S$-transformations: the corresponding hypergeometric functions are related via 
the Gauss relations for the adjacent functions or their iterations.
The action of $S$-transformations on the triples is as follows
\begin{equation}
 \label{eq:equiv1}
\hat\theta_l=\theta_l+n_l,\quad l=0,\,1,\,\infty,\quad 
n_l\in\mathbb{Z},\quad
n_0+n_1+n_\infty=0(\mathrm{mod}\,2).
\end{equation}
Another simple transformation whose action is well known is
the fractional-linear transformations of $\lambda$. The action
of the generating transformations $\lambda\to1/\lambda$ and
$\lambda\to1-\lambda$ on the triples are
\begin{equation}
 \label{eq:equiv2}
(\theta_0,\theta_1,\theta_\infty)\to(\theta_\infty,\theta_1,\theta_0),\quad
\mathrm{and}\quad
(\theta_0,\theta_1,\theta_\infty)\to(\theta_1,\theta_0,\theta_\infty).
\end{equation}
Consider the following transformation of Eq.~(\ref{eq:PHI}), 
$$
\Phi_1(\lambda)=\sigma_1\Phi(\lambda)\sigma_1,\qquad
\sigma_1=\left(\begin{array}{cc}
0&1\\
1&0
\end{array}\right).
$$
This transformation can be interpreted in terms of the formal monodromies
twofold:\\ 
1) as the transformation,
\begin{equation}
 \label{eq:equiv3}
(\theta_0,\theta_1,\theta_\infty)\to(\theta_0,\theta_1,-\theta_\infty),
\end{equation}
which is equivalent to $(\alpha,\beta,\delta)\to(\beta,\alpha,\delta)$;
and 2)
\begin{equation}
 \label{eq:eqiv4} 
(\theta_0,\theta_1,\theta_\infty)\to(-\theta_0,\theta_1,-\theta_\infty)
\end{equation}
or $(\alpha,\beta,\delta-1)\to(\beta-\delta+1,\alpha-\delta+1,1-\delta)$.
Therefore, one can always fix without lost of generality: 
$\theta_0>0$ and $\theta_\infty>0$.
There is one more analogous transformation of Eq.~(\ref{eq:PHI}),
$$
\Phi_1(\lambda)=\left(\begin{array}{cc}
0&\beta_0\\
\alpha_0&0
\end{array}\right)\Phi(\lambda)
\left(\begin{array}{cc}
0&1/\alpha_0\\
1/\beta_0&0
\end{array}\right),\qquad
\alpha_0=\alpha(\alpha-\delta+1),\;
\beta_0=\beta(\beta-\delta+1).
$$ 
The latter transformation results in the following transformation of
the formal monodromy,
\begin{equation}
 \label{eq:equiv5}
(\theta_0,\theta_1,\theta_\infty)\to(\theta_0,-\theta_1,-\theta_\infty),
\end{equation}
or $(\alpha,\beta,\delta)\to(-\alpha+\delta-1,-\beta+\delta-1,\delta)$.
Therefore, we can restrict our consideration only to the case of
positive formal monodromies, $\theta_0>0$, $\theta_1>0$, and $\theta_\infty>0$.
 
{\it We call two $RS$-transformations of Eq.~{\rm(\ref{eq:PHI})} equivalent 
if they are related with the equivalent $R$-transformations and  
their corresponding {\rm(}modulo Eq.~{\rm(\ref{eq:equiv2}))} formal monodromies 
are related via the commutative group of transformations given by
Eqs.~{\rm(\ref{eq:equiv1}), (\ref{eq:equiv3})--(\ref{eq:equiv5})} and the 
identical transformation}.
\begin{remark}
{(\bf Notational comment)} To simplify notation in Subsections of the
following Sections we use the same notation for similar quantities  
corresponding to different transformations. This should not confuse
the reader, since the values of these quantities are
``locally defined'' and valid only inside of the corresponding subsection.
At the same time we have global notation valid through the whole paper,
this is the notation introduced for the function 
$\Phi(\lambda;\theta_0,\theta_1,\theta_\infty)$, in Section {\rm\ref{sec:hyper}}
in Eq.~{\rm(\ref{eq:PHI_HYPER})} and below to the end of the section.
\end{remark}
\section{Quadratic and Cubic Transformations}
 \label{sec:qucu}
Now we are ready to discuss briefly the seed quadratic and 
cubic transformations for the hypergeometric function or,
by other words, the equivalence classes of $RS$-transforma\-tions of 
Eq.~(\ref{eq:PHI}) of the ranks $2$ and $3$.
\subsection{Quadratic transformations}
It is easy to prove that there is only one class of $R$-transformations
which generates $RS$-transformations of the rank $2$. That is,
$R(2|1+1|2)$. This transformation is just 
\begin{equation}
 \label{eq:lambda2}
\lambda=-4\lambda_1(\lambda_1-1)\qquad\lambda-1=-(2\lambda_1-1)^2.
\end{equation}
To this $R$-transformation corresponds only one class of the 
$RS$-transformations, that is $RS_3^2(1+1|2|2)$. In terms of the
formal monodromies it reads as follows,
\begin{equation}
 \label{eq:quadratic}
(\theta_0,\frac12,\theta_\infty)\to(\theta_0,\theta_0,2\theta_\infty-1).
\end{equation}
The complete description of $RS_3(1+1|2|2)$ is as follows:
\begin{equation}
 \label{eq:abd2}
\alpha=\frac{\theta_\infty}2-\frac14-\frac{\theta_0}2,\qquad
\beta=-\frac{\theta_\infty}2-\frac14-\frac{\theta_0}2,\qquad
\delta=1-\theta_0.
\end{equation}
Corresponding matrices $A$ and $B$ in Eq.~(\ref{eq:PHI}) read,
\begin{equation}
 \label{eq:AB22}
A=\frac1{16\theta_\infty}\left(\begin{array}{cc}
1-4\theta_0^2-4\theta_\infty^2&4\theta_0^2-(1+2\theta_\infty)^2\\
(2\theta_\infty-1)^2-4\theta_0^2&4\theta_0^2+4\theta_\infty^2-1
\end{array}\right),\qquad
B=-A-\frac12\theta_\infty\sigma_3.
\end{equation}
The $RS$-transformation can be written as follows,
\begin{equation}
 \label{eq:21}
\Phi\left(\lambda;\theta_0,\frac12,\theta_\infty\right)=
S(\lambda_1)p^{\frac{\sigma_3}2}
\Phi_1\left(\lambda_1;\theta_0,\theta_0,2\theta_\infty-1\right)V,
\end{equation}
where $\lambda$ is given by Eq.~(\ref{eq:lambda2});
$\lambda_1$ belongs to a neighborhood of $0$; 
$S(\lambda_1)$ is the matrix defining Schlesinger transformations
removing an apparent singularity at $\lambda_1=1/2$,
$$
S(\lambda_1)=\sqrt{\lambda_1-\frac12}\left(\begin{array}{cc}
0&0\\
0&1
\end{array}\right)+
\frac1{\sqrt{\lambda_1-\frac12}}
\left(\begin{array}{cc}
\frac{(2\theta_\infty+1)^2-4\theta_0^2}{(2\theta_\infty-1)^2-4\theta_0^2}&0\\
1&0
\end{array}\right);
$$
$p=\frac1{8\theta_\infty}(2\theta_0+1-2\theta_\infty)$;
the function $\Phi_1(\lambda_1;\theta_0,\theta_0,2\theta_\infty-1)$ solves 
Eq.~(\ref{eq:PHI}) with $\lambda\to\lambda_1$ and the matrices 
$A$ and $B$ changed respectively by $A_1$ and $B_1$,
\begin{equation}
 \label{eq:AB2new}
A_1=\frac14\left(\begin{array}{cc}
1-2\theta_\infty&1-2\theta_\infty-2\theta_0\\
-1+2\theta_\infty-2\theta_0&2\theta_\infty-1
\end{array}\right),\qquad
B_1=-A_1-\frac{2\theta_\infty-1}2\sigma_3,
\end{equation}
and is given by Eq.~(\ref{eq:PHI_HYPER}) with 
$(\lambda,\alpha,\beta,\delta)$ changed to
$(\lambda_1,\alpha_1,\beta_1,\delta_1)$, where
\begin{equation}
 \label{eq:abd2new}
\alpha_1=\theta_\infty-\frac12-\theta_0,\qquad
\beta_1=-\theta_\infty+\frac12-\theta_0,\qquad
\delta_1=1-\theta_0;
\end{equation}
finally, the matrix $V$, which is independent of $\lambda_1$, can be
calculated by means of asymptotic expansions (\ref{as:0})--(\ref{as:infty}). 
Actually, this matrix can be calculated by comparison of the asymptotic 
expansions
of the left- and right-hand sides of Eq.~(\ref{eq:21}) at any of the points
$\lambda_1=0$, $1$, and $\infty$. For example, from the comparison of 
asymptotics at
$\lambda_1=\infty$ one immediately concludes that $V$ is a diagonal matrix.
However, to calculate it explicitly one needs few further terms in asymptotics
(\ref{as:infty}) which are not given here. Instead of calculation of the absent 
terms
one can compare corresponding asymptotics at $\lambda_1=0$. By doing so we find
for the matrix $V$ the following representation,
$$
V=C_0^{-1}\left(\alpha_1,\beta_1,\delta_1\right)D_0^{-1}2^{\theta_0\sigma_3}
C_0\left(\alpha,\beta,\delta\right)=
\left(\begin{array}{cc}
v&0\\
0&\det V/v
\end{array}\right),
$$
where (also diagonal!) matrix 
$$
D_0=G_0^{-1}(\alpha,\beta,\delta)S(0)p^{\frac{\sigma_3}2}G_0(\alpha_1,\beta_1,\d
elta_1)=
\sqrt{\frac{\theta_\infty}{2\theta_\infty-1-2\theta_0}}
\left(\begin{array}{cc}
\frac{2\theta_\infty+1+2\theta_0}{2\theta_\infty-1+2\theta_0}&0\\
0&\frac{2\theta_\infty+1-2\theta_0}{2\theta_\infty-1}
\end{array}\right)
$$
and
$$
v=\det V\,2^{\frac32-\theta_\infty}e^{\frac{i\pi}4(2\theta_\infty-1+2\theta_0)}
\sqrt{\frac{\theta_\infty}{2\theta_\infty-1-2\theta_0}},\quad
\det V=\frac{(2\theta_\infty-1)^2-4\theta_0^2}{(2\theta_\infty+1)^2-4\theta_0^2}
e^{-\pi i\theta_0}.
$$
In the above formulas the non-diagonal matrices $C_0(\cdot,\cdot,\cdot)$ and 
$G_0(\cdot,\cdot,\cdot)$ are defined at the end of Section \ref{sec:hyper}. They 
depend on the parameters given by Eqs.~(\ref{eq:abd2new}) and (\ref{eq:abd2}).

This transformation is the seed one for the whole bunch of the quadratic
transformations for the hypergeometric function.

\subsection{Cubic transformations}
A simple examination of possible $R$-transformations which generate 
$RS$-transformations of the rank $3$ shows that there exist
only three such transformations that is,
\begin{enumerate}
\item
$R(2+1|2+1|2+1):\quad\lambda=
\rho\lambda_1\frac{(\lambda_1-a)^2}{\lambda_1-b)^2},\quad
\lambda-1=
\rho(\lambda_1-1)\frac{(\lambda_1-c)^2}{\lambda_1-b)^2},$
$$
a=\frac{(3\sqrt{\rho}-1)(1+\sqrt{\rho})}{4\rho},\qquad
b=\frac{(1+\sqrt{\rho})^2}{4\sqrt{\rho}},\qquad
c=\frac{(1+\sqrt{\rho})^2}{4\rho},\qquad\rho\in\mathbb{C};
$$
\item
$R(3|3|1+1+1):\quad\lambda=-\frac{i}{3\sqrt{3}}
\frac{\left(\lambda_1-1/2-i\sqrt{3}/2\right)^3}{\lambda_1(\lambda_1-1)},\quad
\lambda-1=-\frac{i}{3\sqrt{3}}
\frac{\left(\lambda_1-1/2+i\sqrt{3}/2\right)^3}{\lambda_1(\lambda_1-1)};$
\item
$R(2+1|2+1|3):\quad\lambda=16\lambda_1\left(\lambda_1-\frac34\right)^2,\quad
\lambda-1=16(\lambda_1-1)\left(\lambda_1-\frac14\right)^2.$
\end{enumerate}
These $R$-transformations generates the following $RS$ ones:
\begin{enumerate}
\item
 \label{rs3:1}
$RS_3^2(2+1|2+1|2+1):\quad(\frac12,\frac12,\frac12)\to(\frac12,\frac12,\frac12);
$
\item
 \label{rs3:2}
$RS_3^2(3|3|1+1+1):\quad
(\frac13,\frac13,\theta_\infty)\to(\theta_\infty,\theta_\infty,\theta_\infty);$
\item
\begin{enumerate}
\item
 \label{rs3:3a}
$RS_3^2\left(
\begin{tabular}{c|c|c}
1/2&$\theta_1$&1/3\\
2+1&2+1&3\\
\end{tabular}\right):\quad
(\frac12,\theta_1,\frac13)\to(\frac12,\theta_1,2\theta_1);$
\item
 \label{rs3:3b}
$RS_3^2\left(
\begin{tabular}{c|c|c}
1/2&1/2&$\theta_\infty$\\
2+1&2+1&3\\
\end{tabular}\right):\quad
(\frac12,\frac12,\theta_\infty)\to(\frac12,\frac12,3\theta_\infty).$
\end{enumerate}
The cases \ref{rs3:1} and \ref{rs3:3b} are not interesting since they are 
related
with the auto-transformations for the hypergeometric function 
$F(\frac{\theta_\infty}2-\frac12,\frac{\theta_\infty}2;\theta_\infty;\lambda)$
and its adjacent ones which are known \cite{BE} as rather simple functions,
$$
F(\frac{\theta_\infty}2-\frac12,\frac{\theta_\infty}2;\theta_\infty;\lambda)=
\left(\frac12+\frac12\sqrt{1-\lambda}\right)^{1-\theta_\infty}.
$$
\end{enumerate}
The $RS$-transformations described in the items 1-3 seems to be the seed ones
for the whole bunch of the cubic transformations for the hypergeometric 
function.
We omit exact formulas for the action of these $RS$-transformations 
on the function $\Phi(\lambda)$, however,
it is interesting to notice that this action is defined via
the formulas quite similar to the one for the quadratic case (\ref{eq:21}), with
only one, so-called elementary, $S$-transformation, i.e., one can say that the
cubic transformations have the same type of complexity as the quadratic. 
This, possibly, explains why quadratic and cubic 
transformations were known for already more than 120 years, whilst the higher 
order
ones the reader will find only in the following Sections.

\section{$RS_3^2(5|2+2+1|3+1+1)$}
 \label{sec:3}
As it is explained in the previous sections the first step in construction of 
any $RS$-transformation for Eq.~(\ref{eq:PHI}) is to build an appropriate 
rational 
transformation of the variable $\lambda$.
\subsection{$R(5|2+2+1|3+1+1)$}
 \label{subsec:41}
The rational transformation of $\lambda$, which we
denote as $R(5|2+2+1|3+1+1)$ reads, 
\begin{equation}
 \label{eq:lambda}
\lambda=\frac{(\lambda_1-a)^5}
{\lambda_1(\lambda_1-1)(\lambda_1-b)^3},\qquad
\lambda-1=\kappa\frac{(\lambda_1-d)^2(\lambda_1-c)^2}
{\lambda_1(\lambda_1-1)(\lambda_1-b)^3},
\end{equation}
where
\begin{eqnarray*}
a=-\frac{961}{135}+\frac{176}{27}\zeta
+\frac{121}{50}\zeta^2+\frac{11}{50}\zeta^3,&&
b=\frac{1817}{81}-\frac{1520}{81}\zeta
-\frac{209}{30}\zeta^2-\frac{19}{30}\zeta^3,\\
d=\frac{189}{85}-\frac{848}{425}\zeta-
\frac{567}{850}\zeta^2-\frac{249}{4250}\zeta^3,&&
c=\frac{66919}{20655}-\frac{188336}{103275}\zeta-
\frac{5743}{7650}\zeta^2-\frac{2689}{38250}\zeta^3,
\end{eqnarray*}
$$
\kappa=\frac{935}9-\frac{800}9\zeta-33\zeta^2-3\zeta^3,
$$
and $\zeta$ is an arbitrary root of the following equation,
\begin{equation}
 \label{eq:zeta}
27\zeta^4+270\zeta^3+530\zeta^2-1600\zeta+800=0.
\end{equation}
Solution of Eq.~(\ref{eq:zeta}) are as follows,
$$
\zeta=-\frac52+\varepsilon_1\frac{35}{18}\sqrt{3}+
i\left(\varepsilon_2\frac5{18}\sqrt{15}+\varepsilon_3\frac12\sqrt{5}\right),
\varepsilon_k=\pm1,\quad k=1,2,3,\quad 
\varepsilon_1\varepsilon_2\varepsilon_3=-1.
$$
For example, in the case $\varepsilon_1=\varepsilon_2=+1$ and 
$\varepsilon_3=-1$ one finds,
\begin{eqnarray*}
a&=&\frac12-i\frac{11}{90}\sqrt{15},\qquad
b=\frac12+i\frac{19}{54}\sqrt{15},\qquad\kappa=i\frac53\sqrt{15},\\
d&=&\frac12-\frac{128}{405}\sqrt{3}+i\frac{29}{810}\sqrt{15},\quad
c=\frac12+\frac{128}{405}\sqrt{3}+i\frac{29}{810}\sqrt{15}.
\end{eqnarray*}

According to what is written in Section \ref{sec:2} we can associate
with this $R$-transformation two different classes of the $RS$-transformations
related with the following choice of the $\theta$-triples:
$$
\left(\frac 15,\frac12,\frac 13\right),\quad\mathrm{and}\quad
\left(\frac 25,\frac12,\frac 13\right).
$$
We consider them in detail in the following subsections.
\par
\begin{flushleft}{\bf 5.2}$\quad\;RS_3^2\left(
\begin{tabular}{c|c|c}
1/5&-1/2&1/3\\
5&2+2+1&3+1+1\\
\end{tabular}\right)$
\end{flushleft}
\par\bigskip
The choice of the parameters corresponding to this transformation is
as follows,
$$
\alpha=\frac{19}{60},\qquad\beta=-\frac1{60},\qquad\delta=\frac45.
$$
Corresponding matrices $A$ and $B$ in Eq.~(\ref{eq:PHI}) read,
\begin{equation}
 \label{eq:AB33}
A=\frac1{1200}\left(\begin{array}{cc}
89&11\\
589&-89
\end{array}\right),\qquad
B=\frac1{1200}\left(\begin{array}{cc}
-289&-11\\
-589&289
\end{array}\right).
\end{equation}
The $RS$-transformation can be written as follows,
\begin{equation}
 \label{eq:51}
\Phi\left(\lambda;\frac15,-\frac12,\frac13\right)=
S_1(\lambda_1)S_2(\lambda_1)Q_Bp^{\frac{\sigma_3}2}
\Phi_1\left(\lambda_1;\frac13,\frac13,\frac12\right)V,
\end{equation}
where $\lambda$ is given by the rational transformation
(\ref{eq:lambda}); $\lambda_1$ belongs to a neighborhood of
$0$, which does not contain the points $a$, $b$, $c$, $d$, and $1$;
$S_1(\lambda_1)$ and $S_2(\lambda_1)$
are the matrices defining Schlesinger transformations
removing apparent singularities at the points
$d$, $a$ and $b$, $c$ respectively,
\begin{eqnarray*}
S_1(\lambda_1)=\mu_1 J_{da}+1/\mu_1 J_{ad},&&
\mu_1=\sqrt{\frac{\lambda_1-d}{\lambda_1-a}},\\  
S_2(\lambda_1)=\mu_2 J_{bc}+1/\mu_2 J_{cb},&&
\mu_2=\sqrt{\frac{\lambda_1-b}{\lambda_1-c}},  
\end{eqnarray*}
\begin{eqnarray*}
J_{ad}&=&\frac1{20}\left(\begin{array}{cc}
19&1\\
19&1
\end{array}\right),\qquad
J_{da}=\frac1{20}\left(\begin{array}{cc}
1&-1\\
-19&19
\end{array}\right),\\
J_{bc}&=&\frac1{20(a-b)}\left(\begin{array}{cc}
19a+d-20b&a-d\\
19a+d-20b&a-d
\end{array}\right),\qquad
J_{cb}=\left(\begin{array}{cc}
1&0\\
0&1
\end{array}\right)-J_{bc},\\
J_{bc}[1,2]&=&1-J_{bc}[1,1]=
\frac{87}{340}-\frac{693}{3400}\zeta-\frac{261}{3400}\zeta^2-
\frac{243}{34000}\zeta^3,
\end{eqnarray*}
for the choice of $\zeta$ given at the end of Subsection \ref{subsec:41}
$J_{bc}[1,2]=1/60+i\sqrt{5}/150$;
the matrix 
$$
Q_B=\frac{\sqrt{66}}{60}\left(\begin{array}{cc}
1&1\\
-\frac{589}{11}&1
\end{array}\right),
$$
it transforms matrix $B$ to the diagonal form, $Q_B^{-1}BQ_B=\frac14\sigma_3$;
the number $p=\frac{2057}{945}-\frac{352}{189}\zeta-\frac{121}{175}\zeta^2-
\frac{11}{175}\zeta^3$, in particular, for the choice of the root $\zeta$ 
considered at the end of Subsection \ref{subsec:41} 
$p=\frac{i11}{21\sqrt{15}}$;
the function $\Phi_1(\lambda_1;\frac13,\frac13,\frac12)$ solves 
Eq.~(\ref{eq:PHI}) with $\lambda\to\lambda_1$ and the matrices 
$A$ and $B$ changed respectively by $A_1$ and $B_1$,
\begin{equation}
 \label{eq:ABnew}
A_1=\frac1{24}\left(\begin{array}{cc}
-3&-7\\
-1&3
\end{array}\right),\qquad
B_1=\frac1{24}\left(\begin{array}{cc}
-3&7\\
1&3
\end{array}\right),
\end{equation}
and is given by Eq.~(\ref{eq:PHI_HYPER}) with 
$(\lambda,\alpha,\beta,\delta)$ changed to
$(\lambda_1,\alpha_1,\beta_1,\delta_1)$, where
\begin{equation}
 \label{eq:abdnew}
\alpha_1=-\frac1{12},\qquad\beta_1=-\frac7{12},\qquad\delta_1=\frac23;
\end{equation}
finally, the matrix $V$, which is independent of $\lambda_1$, can be
calculated by means of asymptotic expansions (\ref{as:0})--(\ref{as:infty}). 
For example, taking the limit $\lambda_1\to0$ in
Eq.~(\ref{eq:51}) and using asymptotics (\ref{as:0}) and (\ref{as:infty}) one 
finds,
$$
V=C_0^{-1}\left(-1/12,-7/12,2/3\right)D_0^{-1}
q^{\frac{\sigma_3}6},
$$
where a diagonal (!) matrix 
$$
D_0=S_1(0)S_2(0)Q_Bp^{\frac{\sigma_3}2}
G_0\left(-1/12,-7/12,2/3\right)=
\left(\begin{array}{cc}
2/\sqrt{-14}&0\\
0&\sqrt{-14}/3
\end{array}\right)
$$
and 
\begin{equation}
 \label{eq:q}
q=-\frac{b^3}{a^5}=-300+250\zeta+\frac{1485}{16}\zeta^2+\frac{135}{16}\zeta^3.
\end{equation}
For the choice of the root $\zeta$ considered at the end of Subsection 
\ref{subsec:41}, $q=-\frac{125}{16}-i\frac{75}{16}\sqrt{15}$.
\begin{equation}
 \label{eq:C01}
C_0^{-1}\left(-1/12,-7/12,2/3\right)=
\left(\begin{array}{cc}
\frac{\sqrt{\pi}(2-\sqrt{3}+i)\Gamma(\frac{11}{12})}
{\sqrt{3}\Gamma(\frac23)\Gamma(\frac34)}&
\frac{7(2+\sqrt{3}+i)\Gamma(\frac7{12})\Gamma(\frac23)}{24\sqrt{\pi}
\Gamma(\frac34)}\\
-\frac{2(\sqrt{3}+1-i(\sqrt{3}-1))\Gamma(\frac34)}
{7\sqrt{3\pi}\Gamma(\frac23)\Gamma(\frac7{12})}&
\frac{(\sqrt{3}-1-i(\sqrt{3}+1))\Gamma(\frac23)\Gamma(\frac34)}
{12\sqrt{\pi}\Gamma(\frac{11}{12})}
\end{array}\right)
\end{equation}

$RS$-transformation found in this subsection connects hypergeometric functions
corresponding to the second and sixth cases of the Schwartz table.
\par
\begin{flushleft}{\bf 5.3}$\quad\;RS_3^2\left(
\begin{tabular}{c|c|c}
-2/5&-1/2&1/3\\
5&2+2+1&3+1+1\\
\end{tabular}\right)$
\end{flushleft}
\par\bigskip
The choice of the parameters in Eq.~(\ref{eq:PHI}) is as follows, 
$$
\alpha=\frac{37}{60},\qquad\beta=\frac{17}{60},\qquad\delta=\frac75.
$$
Corresponding matrices $A$ and $B$ read,
$$
A=\frac1{1200}\left(\begin{array}{cc}
-19&119\\
481&19
\end{array}\right),\qquad
B=\frac1{1200}\left(\begin{array}{cc}
-181&-119\\
-481&181
\end{array}\right).
$$
The $RS$-transformation can be written as follows,
\begin{equation}
 \label{eq:52}
\Phi\left(\lambda;-\frac25,-\frac12,\frac13\right)=
S_1(\lambda_1)S_2(\lambda_1)S_3(\lambda_1)Q_\infty p^{\frac{\sigma_3}2}
\Phi_1\left(\lambda_1;\frac13,\frac13,\frac12\right)V,
\end{equation}
where $\lambda$ is given by the rational transformation
(\ref{eq:lambda}); $\lambda_1$ belongs to a neighborhood of
$0$, which does not contain the points $a$, $b$, $c$, $d$, and $1$;
$S_1(\lambda_1)$, $S_2(\lambda_1)$,
and $S_3(\lambda_1)$
are the matrices defining Schlesinger transformations
removing apparent singularities at the points
$d$ and $b$, $c$, and $a$ respectively,
\begin{eqnarray*}
S_1(\lambda_1)=\mu_1 J_{db}+1/\mu_1 J_{bd},&&
\mu_1=\sqrt{\frac{\lambda_1-d}{\lambda_1-b}},\\  
S_2(\lambda_1)=\mu_2 J_{ac}+1/\mu_2 J_{ca},&&
\mu_2=\sqrt{\frac{\lambda_1-a}{\lambda_1-c}},\\
S_3(\lambda_1)=\mu_3J_{\infty a}+1/\mu_3J_{a\infty},&&
\mu_3=\sqrt{\lambda_1-a}, 
\end{eqnarray*}
\begin{eqnarray*}
J_{bd}&=&\left(\begin{array}{cc}
1&0\\
1&0
\end{array}\right),\qquad
J_{db}=\left(\begin{array}{cc}
0&0\\
-1&1
\end{array}\right),\\
J_{ac}&=&\frac1{20(a-b)}\left(\begin{array}{cc}
13a+7d-20b&7(a-d)\\
13a+7d-20b&7(a-d)
\end{array}\right),\quad
J_{ca}=
\left(\begin{array}{cc}
1&0\\
0&1
\end{array}\right)-J_{ac},\\
J_{ac}[1,2]&=&1-J_{ac}[1,1]=
\frac{609}{340}-\frac{4851}{3400}\zeta-\frac{1827}{3400}\zeta^2-
\frac{1701}{34000}\zeta^3,\\
J_{a\infty}&=&\frac1{300}\left(\begin{array}{cc}
251&49\\
251&49
\end{array}\right),\qquad
J_{\infty a}=\frac1{300}\left(\begin{array}{cc}
49&-49\\
-251&251
\end{array}\right);
\end{eqnarray*}
$$
Q_\infty=\frac1{\sqrt{-\frac{251}{49}-Q}}\left(\begin{array}{cc}
1&1\\
Q&-\frac{251}{49}
\end{array}\right),\;
Q=\frac1{7261}(329161-288000\zeta-106920\zeta^2-9720\zeta^3),
$$
for the value of $\zeta$ chosen at the end of Subsection \ref{subsec:41}
$J_{ac}[1,2]=7/60+i7\sqrt{5}/150$, 
$Q=(-7439+i5400\sqrt{15})/7261$;
the number 
$p=\frac{8353}{2205}-\frac{32}9\zeta-\frac{33}{25}\zeta^2-\frac3{25}\zeta^3$, 
in particular, for the choice of the root $\zeta$ 
considered at the end of Subsection \ref{subsec:41}, 
$p=-\frac{18}{49}+i\frac1{\sqrt{15}}$;
the function $\Phi_1(\lambda_1;\frac13,\frac13,\frac12)$ is exactly
the same as the one in the previous subsection, i.e., it solves 
Eq.~(\ref{eq:PHI}) with $\lambda\to\lambda_1$ and the matrices 
$A$ and $B$ changed respectively by $A_1$ and $B_1$ which are given
in Eq.~(\ref{eq:ABnew}); Eq.~(\ref{eq:PHI_HYPER}) with 
$(\lambda,\alpha,\beta,\delta)$ changed to
$(\lambda_1,\alpha_1,\beta_1,\delta_1)$ given in Eq.~(\ref{eq:abdnew}) 
defines an explicit representation for 
$\Phi_1(\lambda_1;\frac13,\frac13,\frac12)$;
finally, the matrix $V$, which is independent of $\lambda_1$, can be
calculated in the same way as in the previous Subsection,
$$
V=C_0^{-1}\left(-1/12,-7/12,2/3\right)D_{0}^{-1}
q^{\frac{\sigma_3}6},
$$
where the number $q$ and the matrix $C_0^{-1}\left(-1/12,-7/12,2/3\right)$
are defined in Eqs.~(\ref{eq:C01}) and (\ref{eq:q}) respectively, whilst
the diagonal (!) matrix $D_{0}$ is different, 
\begin{eqnarray*}
D_0&=&S_{1}(0)S_{2}(0)S_3(0)Q_\infty p^{\frac{\sigma_3}2}
G_0\left(-1/12,-7/12,2/3\right)=
\left(\begin{array}{cc}
s&0\\
0&\frac2{3s}
\end{array}\right),\\
s&=&\sqrt{-\frac{46}3+\frac{40}3\zeta+\frac{99}{20}\zeta^2+\frac9{20}\zeta^3}.
\end{eqnarray*}
For our choice of $\zeta$ at the end of Subsection \ref{subsec:41}
$s=1/4(\sqrt{10}-i\sqrt{6})$.

$RS$-transformation found in this subsection connects hypergeometric functions
corresponding to the second and fourteen cases of the Schwartz table.
\section{$RS_3^2(5+1|2+2+2|3+2+1)$}
 \label{sec:6}
\subsection{$R(5+1|2+2+2|3+2+1)$}
 \label{subsec:51}
The rational transformation of the variable $\lambda$, which we
denote as $R(5+1|2+2+2|3+2+1)$ reads, 
\begin{equation}
 \label{eq:lambda6}
\lambda=\frac4{27}\frac{\lambda_1(16\lambda_1+9)^5}
{(\lambda_1-1)^2(128\lambda_1-3)^3},\qquad
\lambda-1=\frac{(2048\lambda_1^3-10944\lambda_1^2+2619\lambda_1+27)^2}
{27(\lambda_1-1)^2(128\lambda_1-3)^3}.
\end{equation}
We denote the root and pole of the largest multiplicities, 
five and three respectively, of the $R$-transformation as 
$$
a=-\frac9{16}=-0.5625,\quad b=\frac3{128}=0.0234375.
$$
The double roots of $\lambda-1$ are denoted as $c_1$, $c_2$, and $c_3$.
Their explicit values are
\begin{eqnarray*}
c_1&=&\frac{57}{32}+
\frac{75}{32}\sqrt{2}\cos\left(\frac13\arctan\left(\frac17\right)\right),\\
c_2&=&\frac{57}{32}-
\frac{75}{64}\sqrt{2}\cos\left(\frac13\arctan\left(\frac17\right)\right)-
\frac{75}{64}\sqrt{6}\sin\left(\frac13\arctan\left(\frac17\right)\right),\\
c_3&=&\frac{57}{32}-
\frac{75}{64}\sqrt{2}\cos\left(\frac13\arctan\left(\frac17\right)\right)+
\frac{75}{64}\sqrt{6}\sin\left(\frac13\arctan\left(\frac17\right)\right).
\end{eqnarray*}
The first digits of these numbers in the floating-point representation
are as follows,
$$
c_1=5.0921060625\ldots,\quad
c_2=-0.0098990450\ldots,\quad
c_3=0.2615429824\ldots\quad.
$$

According to Section \ref{sec:2} with this $R$-transformation one 
can associate four different classes of the $RS$-transformations
corresponding to the following $\theta$-triples:
$$
\left(\frac15,\frac12,\frac13\right),\quad
\left(\frac25,\frac12,\frac13\right),\quad
\left(\frac15,\frac12,\frac12\right),\quad
\mathrm{and}\quad
\left(\frac 25,\frac12,\frac 12\right).
$$
These transformations are considered in the following subsections.
\par
\begin{flushleft}{\bf 6.2}$\quad\;RS_3^2\left(
\begin{tabular}{c|c|c}
1/5&-1/2&1/3\\
5+1&2+2+2&3+2+1\\
\end{tabular}\right)$
\end{flushleft}
\par\bigskip
The choice of the parameters which corresponds to this transformation 
is exactly the same as in Subsection 5.2:
$$
\alpha=\frac{19}{60},\qquad\beta=-\frac1{60},\qquad\delta=\frac45.
$$
Clearly that the corresponding matrices $A$ and $B$ in Eq.~(\ref{eq:PHI})
also coincide with those given in Eq.~(\ref{eq:AB33}).
The $RS$-transformation reads,
\begin{equation}
 \label{eq:61}
\Phi\left(\lambda;\frac15,-\frac12,\frac13\right)=
S_1(\lambda_1)S_2(\lambda_1)S_3(\lambda_1)
\Phi_1\left(\lambda_1;-\frac15,-\frac13,\frac13\right)
\left(\frac{27}{2}\right)^{\frac{\sigma_3}3},
\end{equation}
where $\lambda$ is given by the rational transformation
(\ref{eq:lambda6}), $\lambda_1$ belongs to a neighborhood
of $\infty$ which does not contain the points $0$, $1$, $a$, $b$,
$c_1$, $c_2$, and $c_3$; 
$S_1(\lambda_1)$, $S_2(\lambda_1)$,
and $S_3(\lambda_1)$
are the matrices defining Schlesinger transformations
removing apparent singularities at the points
$a$ and $c_1$, $b$ and $c_2$, and $c_3$ respectively,
\begin{eqnarray*}
S_1(\lambda_1)=\mu_1 J_{ac_1}+1/\mu_1 J_{c_1a},&&
\mu_1=\sqrt{\frac{\lambda_1-a}{\lambda_1-c_1}},\\  
S_2(\lambda_1)=\mu_2 J_{bc_2}+1/\mu_2 J_{c_2b},&&
\mu_2=\sqrt{\frac{\lambda_1-b}{\lambda_1-c_2}},\\
S_3(\lambda_1)=\mu_3J_{1c_3}+1/\mu_3J_{c_31},&&
\mu_3=\sqrt{\frac{\lambda_1-1}{\lambda_1-c_3}}, 
\end{eqnarray*}
\begin{eqnarray*}
J_{c_1a}&=&\frac1{20}\left(\begin{array}{cc}
1&-1\\
-19&19
\end{array}\right),\qquad
J_{ac_1}=\frac1{20}\left(\begin{array}{cc}
19&1\\
19&1
\end{array}\right),\\
J_{bc_2}&=&\frac15\left(\begin{array}{cc}
4-{\cal C}&1+{\cal C}\\
4-{\cal C}&1+{\cal C}
\end{array}\right),\quad
J_{c_2b}=\frac15
\left(\begin{array}{cc}
1+{\cal C}&-1-{\cal C}\\
-4+{\cal C}&4-{\cal C}
\end{array}\right),\\
{\cal C}&=&\sqrt{2}\cos\left(\frac13\arctan\left(\frac17\right)\right),
\quad
{\cal S}=\sqrt{2}\sin\left(\frac13\arctan\left(\frac17\right)\right),\\
J_{1c_3}&=&\left(\begin{array}{cc}
x&xy\\
\frac{1-x}y&1-x
\end{array}\right),\qquad
J_{c_31}=\left(\begin{array}{cc}
1-x&-xy\\
-\frac{1-x}y&x
\end{array}\right),\\
x&=&\frac7{10}+\frac{3\sqrt{3}}{10}{\cal CS}+\frac{3\sqrt{6}}{20}{\cal S}-
\frac{3\sqrt{2}}{20}{\cal C}+\frac3{10}{\cal C}^2,\\
y\!&=&\!\!\frac3{\sqrt{2}}\frac{(\sqrt{2}{\cal C}+1)
(12{\cal C}^3+18\sqrt{2}{\cal C}^2-38{\cal C}+5\sqrt{2}-
12\sqrt{3}{\cal S}{\cal C}^2+6\sqrt{6}{\cal S}{\cal C}-8\sqrt{3}{\cal S})}
{(66\sqrt{3}{\cal S}{\cal C}+42\sqrt{6}{\cal S}+125+12\sqrt{2}{\cal C}+
120{\cal C}^2-72\sqrt{2}{\cal C}^3-36{\cal C}^4+36\sqrt{3}{\cal S}{\cal C}^3)},
\end{eqnarray*}
The function $\Phi_1(\lambda_1;-\frac15,-\frac13,\frac13)$ solves 
Eq.~(\ref{eq:PHI}) with $\lambda\to\lambda_1$ and the matrices 
$A$ and $B$ changed respectively by $A_1$ and $B_1$,
\begin{equation}
 \label{eq:ABnew6}
A_1=\frac1{300}\left(\begin{array}{cc}
-9&9\\
91&9
\end{array}\right),\qquad
B_1=\frac1{300}\left(\begin{array}{cc}
-41&-9\\
-91&41
\end{array}\right),
\end{equation}
and is given by Eq.~(\ref{eq:PHI_HYPER}) with 
$(\lambda,\alpha,\beta,\delta)$ changed to
$(\lambda_1,\alpha_1,\beta_1,\delta_1)$, where
\begin{equation}
 \label{eq:abdnew6}
\alpha_1=\frac{13}{30},\qquad\beta_1=\frac1{10},\qquad\delta_1=\frac65.
\end{equation}

Some transformations of the sixth order are compositions
of the quadratic and cubic transformations. Let us address this question
in more detail to check that this does not apply to the transformation
constructed in this subsection.
Let us list corresponding $\theta$-triples which can be obtained
by application of the quadratic and cubic transformations to the
$\theta$-triple $\left(1/2,1/3,1/5\right)$, which is equivalent to the
one considered in this subsection. We don't
indicate below explicitly transformations (\ref{eq:equiv1})--(\ref{eq:equiv5}) 
which, of course, essential at the level of the equations (functions):
\begin{enumerate}
\item
A quadratic transform (\ref{eq:quadratic}),
$$
\left(\frac12,\frac13,\frac15\right)\to\left(\frac15,\frac15,\frac13\right);
$$
\item
A combination of the quadratic transform (\ref{eq:quadratic}) with cubic
(\ref{rs3:2}),
$$
\left(\frac12,\frac13,\frac15\right)\to\left(\frac13,\frac13,\frac13\right)\to
\left(\frac35,\frac35,\frac35\right);
$$
\item
Combinations of the cubic transform (\ref{rs3:3a}) with quadratic ones 
(\ref{eq:quadratic}),
\begin{enumerate}
\item
$$
\left(\frac12,\frac13,\frac15\right)\to\left(\frac12,\frac15,\frac25\right)\to
\left(\frac15,\frac15,\frac15\right);
$$
\item
$$
\left(\frac12,\frac13,\frac15\right)\to\left(\frac12,\frac15,\frac25\right)\to
\left(\frac25,\frac25,\frac35\right).
$$
\end{enumerate}
\end{enumerate}
These transformations connect hypergeometric functions corresponding to
the following cases of the Schwartz table:
$6$ and $8$; $6$, $7$, and $11$; $6$, $9$, and $13$; and $6$, $9$, and $11$,
respectively. $RS$-transformation found in this subsection connects 
hypergeometric 
functions corresponding to the cases $6$ and $12$ of the Schwartz table,
therefore it can't be obtained as a combination of the lower order 
transformations.
\begin{remark}
 \label{rem:schwartz}
For exact correspondence one should notice that our notation
$\theta_0$, $\theta_1$, and $\theta_\infty$ are
related with the variables $\lambda$, $\mu$, and $\nu$ introduced by 
H.~A.~Schwartz
{\rm(}see {\rm\cite{BE})} as follows
$\lambda=\theta_0$, $\nu=\theta_1$, $\mu=\theta_\infty-1$. 
\end{remark}
\par
\begin{flushleft}{\bf 6.3}$\quad\;RS_3^2\left(
\begin{tabular}{c|c|c}
2/5&-1/2&1/3\\
5+1&2+2+2&3+2+1\\
\end{tabular}\right)$
\end{flushleft}
\par\bigskip
The choice of the parameters which correspond to this transformation is
as follows,
$$
\alpha=\frac{13}{60},\qquad\beta=-\frac7{60},\qquad\delta=\frac35.
$$
The corresponding matrices $A$ and $B$ in Eq.~(\ref{eq:PHI}) read,
\begin{equation}
 \label{eq:AB44}
A=\frac1{1200}\left(\begin{array}{cc}
-19&119\\
481&19
\end{array}\right),\qquad
B=\frac1{1200}\left(\begin{array}{cc}
-181&-119\\
-481&181
\end{array}\right).
\end{equation}
The form of the $RS$-transformation is as follows,
\begin{equation}
 \label{eq:62}
\Phi\left(\lambda;\frac25,-\frac12,\frac13\right)=
S_1(\lambda_1)S_2(\lambda_1)S_3(\lambda_1)
\left(\frac{\sqrt{11}}6\right)^{\sigma_3}
\Phi_1\left(\lambda_1;\frac25,\frac23,\frac13\right)
\left(\frac{9\sqrt[3]{4}}{\sqrt{11}}\right)^{\sigma_3},
\end{equation}
where $\lambda$ is given by the rational transformation
(\ref{eq:lambda6}), $\lambda_1$ belongs to a neighborhood
of $\infty$ which does not contain the points $0$, $1$, $a$, $b$,
$c_1$, $c_2$, and $c_3$; 
$S_1(\lambda_1)$, $S_2(\lambda_1)$,
and $S_3(\lambda_1)$
are the matrices defining Schlesinger transformations
removing apparent singularities at the points
$b$ and $c_1$, $c_2$, and $c_3$ respectively,
\begin{eqnarray*}
S_1(\lambda_1)=\mu_1 J_{bc_1}+1/\mu_1 J_{c_1b},&&
\mu_1=\sqrt{\frac{\lambda_1-b}{\lambda_1-c_1}},\\  
S_2(\lambda_1)=\mu_2 J_{ac_2}+1/\mu_2 J_{c_2a},&&
\mu_2=\sqrt{\frac{\lambda_1-a}{\lambda_1-c_2}},\\
S_3(\lambda_1)=\mu_3J_{ac_3}+1/\mu_3J_{c_3a},&&
\mu_3=\sqrt{\frac{\lambda_1-a}{\lambda_1-c_3}}, 
\end{eqnarray*}
\begin{eqnarray*}
J_{c_1b}&=&\left(\begin{array}{cc}
0&0\\
-1&1
\end{array}\right),\qquad
J_{bc_1}=\left(\begin{array}{cc}
1&0\\
1&0
\end{array}\right),\\
J_{ac_2}&=&\frac15\left(\begin{array}{cc}
-2-7{\cal C}&7+7{\cal C}\\
-2-7{\cal C}&7+7{\cal C}
\end{array}\right),\quad
J_{c_2a}=\frac15
\left(\begin{array}{cc}
7+7{\cal C}&-7-7{\cal C}\\
2+7{\cal C}&-2-7{\cal C}
\end{array}\right),\\
{\cal C}&=&\sqrt{2}\cos\left(\frac13\arctan\left(\frac17\right)\right),
\quad
{\cal S}=\sqrt{2}\sin\left(\frac13\arctan\left(\frac17\right)\right),\\
J_{ac_3}&=&\left(\begin{array}{cc}
1-x&x\\
1-x&x
\end{array}\right),\qquad
J_{c_3a}=\left(\begin{array}{cc}
x&-x\\
x-1&1-x
\end{array}\right),
\end{eqnarray*}
$$
x=\frac{7(3+\sqrt{3})}{600}
((20\sqrt{3}-24){\cal C}^2-12{\cal C}{\cal S}
-10\sqrt{3}+12+(7\sqrt{6}-9\sqrt{2}){\cal C}+(\sqrt{6}+3\sqrt{2}){\cal S}),
$$
The function $\Phi_1(\lambda_1;\frac25,\frac23,\frac13)$ solves 
Eq.~(\ref{eq:PHI}) with $\lambda\to\lambda_1$ and the matrices 
$A$ and $B$ changed respectively by $A_1$ and $B_1$,
$$
A_1=\frac1{300}\left(\begin{array}{cc}
39&-189\\
-11&-39
\end{array}\right),\qquad
B_1=\frac1{300}\left(\begin{array}{cc}
-89&189\\
11&89
\end{array}\right),
$$
and is given by Eq.~(\ref{eq:PHI_HYPER}) with 
$(\lambda,\alpha,\beta,\delta)$ changed to
$(\lambda_1,\alpha_1,\beta_1,\delta_1)$, where
$$
\alpha_1=-\frac{11}{30},\qquad\beta_1=-\frac7{10},\qquad\delta_1=\frac35.
$$
As in the previous subsection we present here the list of the
$\theta$-triples which can be obtained
by application of the quadratic and cubic transformations to the equations 
(functions)
corresponding to the $\theta$-triple $\left(1/2,1/3,2/5\right)$ to prove that 
our
transformation is not a composition of the lower order ones. As before we don't
indicate explicitly in the list below transformations 
(\ref{eq:equiv1})--(\ref{eq:equiv5}) and
keep in mind Remark \ref{rem:schwartz}:
\begin{enumerate}
\item
A quadratic transform (\ref{eq:quadratic}),
$$
\left(\frac12,\frac13,\frac25\right)\to\left(\frac25,\frac25,\frac13\right);
$$
\item
A combination of the quadratic transform (\ref{eq:quadratic}) with cubic
(\ref{rs3:2}),
$$
\left(\frac12,\frac13,\frac25\right)\to\left(\frac13,\frac13,\frac15\right)\to
\left(\frac15,\frac15,\frac15\right);
$$
\item
Combinations of the cubic transform (\ref{rs3:3a}) with quadratic ones 
(\ref{eq:quadratic}),
\begin{enumerate}
\item
$$
\left(\frac12,\frac13,\frac25\right)\to\left(\frac12,\frac25,\frac15\right)\to
\left(\frac15,\frac15,\frac15\right);
$$
\item
$$
\left(\frac12,\frac13,\frac25\right)\to\left(\frac12,\frac25,\frac15\right)\to
\left(\frac25,\frac25,\frac35\right).
$$
\end{enumerate}
\end{enumerate}
These transformations connect hypergeometric functions corresponding to
the following cases of the Schwartz table:
$14$ and $15$; $14$, $12$, and $13$; $14$, $9$, and $13$; and $14$, $9$, and 
$11$,
respectively. $RS$-transformation found in this subsection connects 
hypergeometric 
functions corresponding to the cases $14$ and $7$ of the Schwartz table,
therefore it can't be obtained as a combination of the lower order 
transformations.
\par
\begin{flushleft}{\bf 6.4}$\quad\;RS_3^2\left(
\begin{tabular}{c|c|c}
1/5&1/2&1/2\\
5+1&2+2+2&3+2+1\\
\end{tabular}\right)$
\end{flushleft}
\par\bigskip
In this case it is convenient to
rescale $R$-transformation (\ref{eq:lambda6}),
\begin{equation}
 \label{eq:lambda6m}
\lambda=\frac1{64}\frac{\lambda_1(\lambda_1+24)^5}
{(\lambda_1-1)^3(3\lambda_1-128)^2},\qquad
\lambda-1=\frac{(\lambda_1^3-228\lambda_1^2+2328\lambda_1+1024)^2}
{64(\lambda_1-1)^3(3\lambda_1-128)^2}.
\end{equation}
We denote the root of the $R$-transformation of the fifth order, 
and the pole of the second as
\begin{equation}
 \label{eq:ab6new}
a=-24,\qquad b=\frac{128}3=42.(6),
\end{equation}
respectively.
The double roots of $\lambda-1$ are denoted as $c_1$, $c_2$, and $c_3$.
Their explicit values are
\begin{eqnarray}
 \label{eq:c1}
c_1&=&76+
100\sqrt{2}\cos\left(\frac13\arctan\left(\frac17\right)\right),\\
c_2&=&76-
50\sqrt{2}\cos\left(\frac13\arctan\left(\frac17\right)\right)-
50\sqrt{6}\sin\left(\frac13\arctan\left(\frac17\right)\right),\\
c_3&=&76-
50\sqrt{2}\cos\left(\frac13\arctan\left(\frac17\right)\right)+
50\sqrt{6}\sin\left(\frac13\arctan\left(\frac17\right)\right).
\label{eq:c3}
\end{eqnarray}
The first digits of these numbers in the floating-point representation
are as follows,
$$
c_1=217.2631920020\ldots,\quad
c_2=-0.4223592539\ldots,\quad
c_3=11.1591672518\ldots\quad.
$$

The choice of the parameters corresponding to this $RS$-transformation is as 
follows,
\begin{equation}
 \label{eq:parameters1}
\alpha=-\frac1{10},\qquad\beta=-\frac35,\qquad\delta=\frac45.
\end{equation}
The corresponding matrices $A$ and $B$ in Eq.~(\ref{eq:PHI}) read,
\begin{equation}
 \label{eq:AB45}
A=\frac1{50}\left(\begin{array}{cc}
-1&-24\\
-1&1
\end{array}\right),\qquad
B=\frac1{100}\left(\begin{array}{cc}
-23&24\\
2&23
\end{array}\right).
\end{equation}
The $RS$-transformation can be written in the following form,
\begin{equation}
 \label{eq:63}
\Phi\left(\lambda;\frac15,\frac12,\frac12\right)=
S_1(\lambda_1)S_2(\lambda_1)S_3(\lambda_1)
\left(i2\sqrt{6}\right)^{\sigma_3}
\Phi_1\left(\lambda_1;\frac15,\frac12,\frac12\right)
\left(-i2\sqrt{6}\right)^{\sigma_3},
\end{equation}
where $\lambda$ is given by the rational transformation
(\ref{eq:lambda6m}), $\lambda_1$ belongs to a neighborhood
of $\infty$ which does not contain the points $0$, $1$, $a$, $b$,
$c_1$, $c_2$, and $c_3$; 
$S_1(\lambda_1)$, $S_2(\lambda_1)$,
and $S_3(\lambda_1)$
are the matrices defining Schlesinger transformations
removing apparent singularities at the points
$a$ and $c_1$, $b$ and $c_2$, and $c_3$ respectively,
\begin{eqnarray*}
S_1(\lambda_1)=\mu_1 J_{ac_1}+1/\mu_1 J_{c_1a},&&
\mu_1=\sqrt{\frac{\lambda_1-a}{\lambda_1-c_1}},\\  
S_2(\lambda_1)=\mu_2 J_{bc_2}+1/\mu_2 J_{c_2b},&&
\mu_2=\sqrt{\frac{\lambda_1-b}{\lambda_1-c_2}},\\
S_3(\lambda_1)=\mu_3J_{1c_3}+1/\mu_3J_{c_31},&&
\mu_3=\sqrt{\frac{\lambda_1-1}{\lambda_1-c_3}}, 
\end{eqnarray*}
\begin{eqnarray*}
J_{c_1a}&=&\frac1{30}\left(\begin{array}{cc}
6&144\\
1&24
\end{array}\right),\qquad
J_{ac_1}=\frac1{30}\left(\begin{array}{cc}
24&-144\\
-1&6
\end{array}\right),\\
J_{bc_2}&=&\left(\begin{array}{cc}
\frac7{10}-\frac3{10}{\cal C}&-\frac{36}5-\frac{36}5{\cal C}\\
&\\
-\frac7{240}+\frac1{80}{\cal C}&\frac3{10}+\frac3{10}{\cal C}
\end{array}\right),\quad
J_{c_2b}=\left(\begin{array}{cc}
\frac3{10}+\frac3{10}{\cal C}&\frac{36}5+\frac{36}5{\cal C}\\
&\\
\frac7{240}-\frac1{80}{\cal C}&\frac7{10}-\frac3{10}{\cal C}
\end{array}\right),\\
{\cal C}&=&\sqrt{2}\cos\left(\frac13\arctan\left(\frac17\right)\right),
\quad
{\cal S}=\sqrt{2}\sin\left(\frac13\arctan\left(\frac17\right)\right),\\
J_{1c_3}&=&\left(\begin{array}{cc}
x&xy\\
(1-x)/y&1-x
\end{array}\right),\qquad
J_{c_31}=\left(\begin{array}{cc}
1-x&-xy\\
(x-1)/y&x
\end{array}\right),
\end{eqnarray*}
\begin{eqnarray*}
x&=&-\frac15-\frac35{\cal C}+\frac35\sqrt{3}{\cal S}+
\frac35\sqrt{3}{\cal C}{\cal S}+\frac35{\cal C}^2,\\
y&=&72\frac{({\cal C}+1)
(6{\cal C}^3-6\sqrt{3}{\cal S}{\cal C}^2+18{\cal C}^2+
6\sqrt{3}{\cal C}{\cal S}-23{\cal C}+7\sqrt{3}{\cal S}-20)}
{18{\cal C}^4-18\sqrt{3}{\cal C}^3{\cal S}+72{\cal C}^3-30{\cal C}^2
+24\sqrt{3}{\cal C}{\cal S}+6\sqrt{3}{\cal S}-114{\cal C}-55}.
\end{eqnarray*}
The function $\Phi_1(\lambda_1;\frac15,\frac12,\frac12)$ solves 
Eq.~(\ref{eq:PHI}) with $\lambda\to\lambda_1$ and the same matrices 
$A$ and $B$ as for the function $\Phi(\lambda;\frac15,\frac12,\frac12)$ 
(see Eq.~(\ref{eq:AB45})) and is given by Eq.~(\ref{eq:PHI_HYPER}) with 
$\lambda$ changed to $\lambda_1$, whilst the parameters
$\alpha$, $\beta$, and $\delta$ are the same as before the transformation
and given by Eq.~(\ref{eq:parameters1}).
This transformation is related with the hypergeometric function,
$F\left(-1/10,1/10;1/2;1/\lambda\right)$. Its trigonometric form is well known
\cite{B,BE}, $F\left(-1/10,1/10;1/2;\sin^2t\right)=\cos\left(\frac15t\right)$.
The algebraic form of this function is as follows,
$$
F\left(-\frac1{10},\frac1{10};\frac12;\frac1{\lambda}\right)=\frac12
\left(\sqrt{-\frac1{\lambda}}+\sqrt{1-\frac1{\lambda}}\right)^{\frac15}+
\frac{\left(\left(2-\frac4\lambda\right)
\left(\sqrt{-\frac1{\lambda}}+\sqrt{1-\frac1{\lambda}}\right)^2-
1\right)^{\frac15}}
{2\left(\sqrt{-\frac1{\lambda}}+\sqrt{1-\frac1{\lambda}}\right)}.
$$
The formula is valid in a proper neighborhood of $\lambda=\infty$.

It is worth to notice, that there are higher-order transformations
related with this function which can be obtained as iterations of
quadratic and cubic transformations (as in the previous subsections
we don't indicate explicitly transformations 
(\ref{eq:equiv1})--(\ref{eq:equiv5})):
\begin{enumerate}
\item
A fourth order transform, as an iteration of two quadratic transformations
(\ref{eq:quadratic});
\item
 \label{tr:six}
A sixth order transform, as the successive (in arbitrary order) quadratic 
(\ref{eq:quadratic}) and cubic (\ref{rs3:3b}) transforms;
\item
A ninth order transform, as an iteration of the cubic (\ref{rs3:3b}) transforms.
\end{enumerate} 
All these three cases, in terms of the $\theta$-triples read identically,
$$
\left(\frac12,\frac12,\frac15\right)\to\left(\frac12,\frac12,\frac35\right)\to
\left(\frac12,\frac12,\frac15\right),
$$
although in terms of the hypergeometric function they are very different.
A comparison of the $R$-transforms, which correspond to the sixth order 
transformations
obtained in this subsection, and the one described in item \ref{tr:six} shows 
that
it is two different transformations of the six order. All these 
auto-transformations 
corresponds to the first case of the Schwartz table.

A situation with $RS$-transformations corresponding to the triple 
$\left(2/5,1/2,1/2\right)$ is absolutely parallel with the one considered in 
this
subsection: corresponding hypergeometric function,
$F\left(-1/5,1/5;1/2;\sin^2t\right)=\cos\left(\frac25t\right)$, is also a
superposition of elementary functions, there are auto-transformations of the
fourth and sixth order which are superposition of the quadratic and cubic 
transformations. Moreover the hypergeometric functions corresponding
to the triples $\left(2/5,1/2,1/2\right)$ and $\left(1/5,1/2,1/2\right)$ 
are related by a quadratic transformation. We omit here explicit formulas 
for the corresponding auto-$RS$-transformation related with the
$R$-transform (\ref{eq:lambda6}), although it is new it does not seem to
add much knowledge to the subject.
\section{Concluding remarks}
 \label{sec:8}
As it was mentioned above most of the higher order transformations relate
hypergeometric functions corresponding to the parameters from 
the Schwartz table. At the same time it is clear that there are higher order
transformations for the hypergeometric functions whose parameters are different
from the ones included in the Schwartz table. The quadratic and cubic 
transformations
(see Section \ref{sec:qucu}), which contain an arbitrary parameter, are the most 
known 
of them. There are also transformations of the fourth, fifth, and sixth order 
with
a parameter which also run out of the Schwartz table. For example, combinations
of the cubic and quadratic transformations considered in Section \ref{sec:qucu}
give sixth order transformations which relate hypergeometric functions 
corresponding to the following $\theta$ parameters:
\begin{eqnarray*}
\left(\frac12,\frac13,\theta_\infty\right)
&\longrightarrow&\left(\theta_\infty,\theta_\infty,4\theta_\infty-1\right),\\
\left(\frac12,\frac13,\theta_\infty\right)
&\longrightarrow&\left(2\theta_\infty,2\theta_\infty,2\theta_\infty-1\right).
\end{eqnarray*}
Here we would like to present interesting $RS$-transformations of
the rank $8$ which are based on the following $R$-transformation,
\begin{equation}
 \label{eq:lambda8}
\lambda=\frac{\rho\lambda_1(\lambda_1-a)^7}
{(\lambda_1-1)(\lambda_1-c_1)^3(\lambda_1-c_2)^3},\;\;
\lambda-1=\frac{\rho(\lambda_1-b_1)^2(\lambda_1-b2)^2(\lambda_1-b3)^2(\lambda_1-
b4)^2}
{(\lambda_1-1)(\lambda_1-c_1)^3(\lambda_1-c_2)^3}.
\end{equation}  
We were not able to find the numbers $\rho$, $a$, $c_1$, $c_2$, $b_1$, $b_2$, 
$b_3$,
and $b_4$ explicitly. However, an application of the Groebner package on MAPLE 6 
code
shows that transformation (\ref{eq:lambda8}) really exists. Moreover, we were 
able
to calculate this transformation numerically. The following list presents first
20 correct digits after the decimal point in the floating-point representation 
of 
the numbers defining transformation (\ref{eq:lambda8}):
\begin{eqnarray*}
\rho&=&1/112-i0.04639421805988064179\dots,\\
a&=&0.27551020408163265306\ldots-i0.68928552546108382089\ldots,\\
c_1&=&0.00025904071151396454\ldots+i0.02275992796424688660\ldots,\\
c_2&=&0.65407769398236358647\ldots+i0.93826316041899497905\ldots,\\
b_1&=&0.00004664085318617660\ldots-i0.00965813289425897678\ldots,\\
b_2&=&0.03098261455781643319\ldots+i0.24699252359291090010\ldots,\\
b_3&=&1.01560634234261281988\ldots+i0.12947315141387009072\ldots,\\
b_4&=&1.91765011653209885603\ldots+i7.61299796418694837399\ldots.\\
\end{eqnarray*} 
As far as $R$-transformation is known it is straightforward to write
an associated $RS$-transformations. In this case, we omit here this formulas,
since we can write them only in terms of unknown numbers $\rho$, $a$, $c_1$, 
$c_2$, $b_1$, $b_2$, $b_3$, which can be defined as a solution of a system of 
polynomial 
equations. In such form the formulas for the $S$-transformations look quite
cumbersome. We can actually associate three different classes of 
$RS$-transformations with
the $R$-transform (\ref{eq:lambda8}). These classes are defined by the
$RS$-transformations whose action on the associated 
$\theta$-triples read:
$$
\left(\frac12,\frac13,\frac17\right)\to\left(\frac13,\frac23,\frac17\right),
\quad
\left(\frac12,\frac13,\frac27\right)\to\left(\frac13,\frac13,\frac27\right),
\quad
\left(\frac12,\frac13,\frac37\right)\to\left(\frac13,\frac23,\frac37\right).
$$
Acting on these $\theta$-triples by other $RS$-transformations of the rank
$\leq6$ we arrive to a cluster (actually three clusters) of the $\theta$-triples
which do not intersect with the Schwartz table and
whose corresponding hypergeometric functions are related via higher order
transformations. 

\end{document}